\documentstyle[epsfig,aps]{revtex}%,twocolumn]{revtex}

   \def\b{\beta }
\def\dg{\dagger}
   
\def\t{\theta }                    
\def\d{\delta }     
\def\o{\omega }   
\def\l{\lambda }

\font\got=eufm10 scaled \magstep1
\font\gotscr=eufm7 scaled \magstep1
\font\gotscrscr=eufm5 scaled \magstep1
\newfam\gotfam
\textfont\gotfam=\got
\scriptfont\gotfam=\gotscr
\scriptscriptfont\gotfam=\gotscrscr
\def\got{\fam\gotfam}

\font\Bbb=msbm10 scaled \magstep1
\font\Bbbscr=msbm7 scaled \magstep1
\font\Bbbscrscr=msbm5 scaled \magstep1
\newfam\Bbbfam
\textfont\Bbbfam=\Bbb
\scriptfont\Bbbfam=\Bbbscr
\scriptscriptfont\Bbbfam=\Bbbscrscr
\def\Bbb{\Bbbfam}

\font\Cal=msbm10 scaled \magstep1
\font\Calscr=msbm7 scaled \magstep1
\font\Calscrscr=msbm5 scaled \magstep1
\newfam\Calfam
\textfont\Calfam=\Cal
\scriptfont\Calfam=\Calscr
\scriptscriptfont\Calfam=\Calscrscr
\def\Cal{\fam\Calfam}
\def\beq{\begin{equation}}
\def\eeq{\end{equation}}
\def\la{\langle}
\def\ra{\rangle}
\def\gappeq{\mathrel{\rlap {\raise.5ex\hbox{$>$}}
{\lower.5ex\hbox{$\sim$}}}}

\def\lappeq{\mathrel{\rlap{\raise.5ex\hbox{$<$}}
{\lower.5ex\hbox{$\sim$}}}}

\newcommand{\lsim}{\raise.3ex\hbox{$<$\kern-.75em\lower1ex\hbox{$\sim$}}}
\newcommand{\gsim}{\raise.3ex\hbox{$>$\kern-.75em\lower1ex\hbox{$\sim$}}}

\begin{document}
\draft

%\hfill {\large Preprint ITP-99-6E

\vskip 1.4cm

\title{\Large {\bf $q$-Boson Approach to Multiparticle Correlations } }

\vskip 2.8cm

\author{
    D.V.~Anchishkin$^{\! }$ {\footnote{E-mail: anch@bitp.kiev.ua }},
    A.M.~Gavrilik$^{\! }$ {\footnote{E-mail: omgavr@bitp.kiev.ua }}
 and N.Z.~Iorgov$^{\! }$ {\footnote{E-mail: mmtpitp@bitp.kiev.ua }}
        }
\vskip 0.90cm
\address{Bogolyubov Institute for Theoretical Physics,
03143 Kiev-143, Ukraine}

\maketitle

%\date{\today}

\vskip 1.8cm

\begin{abstract}
An approach is proposed enabling to effectively describe, for
relativistic heavy-ion collisions, the observed deviation from unity
of the intercept $\lambda$ (measured value corresponding to zero relative
momentum ${\bf p}$ of two registered identical pions or kaons) of
the two-particle correlation function $C(p,K)$.
The approach uses $q$-deformed oscillators and the related picture of
ideal gas of $q$-bosons.
In effect, the intercept $\lambda$ is connected with deformation
parameter $q$.
For a fixed value of $q$, the model predicts specific dependence
of $\lambda$ on pair mean momentum ${\bf K}$ so that,
when $|{\bf K}|\ \gsim\ 500\!-\!600$ MeV/c for pions
or when $|{\bf K}|\ \gsim\ 700\!-\!800$ MeV/c for kaons,
the intercept $\lambda$ tends to a constant which is less than unity
and determined by $q$.
If $q$ is fixed to be the same for pions and kaons,
the intercepts $\lambda_\pi$ and $\lambda_K$ essentially
differ at small mean momenta ${\bf K}$, but tend to be equal
at ${\bf K}$ large enough ($|{\bf K}|\ \gsim\ \! 800$ MeV/c),
where the effect of resonance decays can be neglected. %is washed out.
We argue that it is of basic interest to check in the experiments
on heavy ion collisions:
(i) the exact shape of dependence $\lambda = \lambda({\bf K})$, and
(ii)
whether for $|{\bf K}|\ \gsim\ \! 800$ MeV/c the resulting
$\lambda_\pi$ and $\lambda_K$ indeed coincide.
\end{abstract}

\vskip 0.5cm


%%%%%%%%%%%%%%%%%
%%%%%%%%%%%%%%%%%

\twocolumn

%%%%%%%%%%%%%%%%%%%%%%%%%%%%%%% sec1 %%%%%%%%%%%%%%%%%%%%%%%%%%%%%%%%%%

%\section{Introduction}
%\label{sec1}

%%%%%%%%%%%%%%%%%%%%%%%%%%%%%%%%%%%%%%%%%%%%%%%%%%%%%%%%%%%%%%%%%%%%%%%

Hadron matter under intense conditions of high temperatures and
densities is
extensively studied in relativistic heavy-ion collisions (RHIC).
Models and approaches used to describe
the processes in the reaction region are examined by comparing
predictions of these models
with experimental data on
single-, two- and many-particle momentum spectra.
Two-particle correlations are known to carry
information about the
space-time structure and dynamics of
the emitting source \cite{GKW,boal,heinz99}. Usually,
the study of correlations occurred in RHIC
assumes that
(i) particles are emitted independently (completely chaotic source), and
(ii) finite multiplicity corrections can be neglected.
Then, correlations reflect
the effects from (anti)symmetrization of the amplitude
to detect identical particles with certain momenta, and
the effects due to final state interactions (FSI) of
detected particles between themselves and with the source.
The FSI, known to depend on the structure of emitting source,
as well provide certain information about source dynamics,
see \cite{anch98}.

The correlation function is defined as \cite{boal}
\begin{equation}
C({\bf k}_a,{\bf k}_b)=
\frac{\displaystyle P_2\left({\bf k}_a, {\bf k}_b\right) }
{\displaystyle P_1\left({\bf k}_a\right) \,
P_1\left({\bf k}_b\right) }
\ ,
\label{i1}
\end{equation}
\noindent
$P_1\left({\bf k}\right)$
and
$P_2\left({\bf k}_a, {\bf k}_b\right)$
being single- and two-particle probabilities to registrate particles with
definite momenta.

For identical particles,
the two-particle wave function appears to be a symmetrized
(antisymmetrized) sum of single-particle states (chaoticity
assumption), namely
 \begin{eqnarray}
\psi_{\gamma_a \gamma_b}({\bf x}_a,{\bf x}_b,t) =
&& \frac{1}{\sqrt{2}} \big[ \psi_{\gamma_a}({\bf x}_a,t)\,
          \psi_{\gamma_b}({\bf x}_b,t) \, +
\nonumber \\
  && + \, e^{i \alpha} \psi_{\gamma_a}({\bf x}_b,t)\,
          \psi_{\gamma_b} ({\bf x}_a,t) \big].  \,
 \label{i5}
 \end{eqnarray}
Here $\alpha =0$ ($\alpha =\pi $) for identical bosons (fermions).
The indices $\gamma_a,\gamma_b$ of the 1-particle wave functions
label complete sets of 1-particle quantum numbers.
Below, we refer our consideration of two-particle
correlations to identical bosons (pions, kaons, etc.).

In the absence of FSI, for chaotic source, the correlation function
can be expressed as follows \cite{AGI}:
\begin{eqnarray}
\label{i2}
&& C(p,K)  = \\
&& =1\, + \, \cos{\alpha} \,
\frac{
\left| \int  d^4 X \, e^{i  p\cdot X } S(X,K) \right| ^2 }
{
 \int  d^4 X \, S\Big( X,K+\frac{p}{2} \Big) \,
 \int  d^4 Y \, S\Big( Y,K-\frac{p}{2} \Big)
}
\ ,
\nonumber
\end{eqnarray}
with the 4-momenta $K$ and $p$ defined as
$K=\frac{1}{2} (k_a+k_b) \, , \ \ p=k_a-k_b \, .  $
The source function $S(X,K)$ (single-particle Wigner density) is
defined by the emitted single-particle states $\psi_\gamma (x)$
at freeze-out times \cite{anch98} (for details of the derivation see
Appendix in \cite{AGI}),  namely
\begin{eqnarray}
\label{i3}
&&  S(X,K) =  \\
&& =  \int d^4x\, e^{i K\cdot x}\,
  \sum_{\gamma , \gamma '} \rho_{\gamma \gamma '}\,
  \psi_\gamma \left(X+{\textstyle{x\over 2}}\right) \,
  \psi_{\gamma '}^*\left(X-{\textstyle{x\over 2}}\right)
  \, .
\nonumber
\end{eqnarray}
Using hermiticity of the density matrix $\rho_{\gamma \gamma '}$
one easily shows that $S(X,K)$ is real.
Due to this, at zero relative momentum one has
$C(0,K) = 1 + \cos\alpha $.
{}From the latter relation it is seen that
the boson correlation function %as follows from (\ref{i2})
at $\alpha=0$ should approach the exact value $2$ as
the relative momentum $\to 0$.
But, as observed from the very first experiments
and up to most recent data,
the measured correlation function never attains this value at ${\bf p}=0$.
To remove this discrepancy, the correlation function of identical bosons
is usually  represented in the form
\begin{equation}
C(p,K) = 1\, + \,  \lambda \, f(p,K)
\label{i6}
\end{equation}
with $\lambda $ drawn from a fit to the data,
$\lambda=$ 0.4\, --\, 0.9, $f(p,K)$ commonly taken as Gaussian,
so that $f({\bf p}=0,K)=1$.
The deviation of $\lambda $ from unity in RHIC can be interpreted by
production of secondary pions from resonance decays
going outside the fireball.
Presence of long-lived resonances results in an increase of measured
source size and life-times \cite{heiselberg96,heinz96}.

%%%%%%%%%%%%%%%%%%%%%%%%%%%%%%%%%%%%%%%%%%%%
Trying to explain experimental data, with Eq.~(\ref{i2}) in mind
(admitting that $\alpha$, besides $0$ or $\pi$, can as well
take other values)
it is natural to relate the parameter $\lambda $ with the effective
angle $\alpha $ and to get {\it the reduction factor}
by  means of $\cos{\alpha}$.
The correlation function (\ref{i2}), which is measurable quantity,
possesses the obvious symmetry $\alpha \to -\alpha$, hence there is
no contradiction in taking the wave function in the form (\ref{i5}).
One can argue that the two-particle wave function  of a boson pair
released from a dense and hot environment effectively acquires
an additional phase. The drawn phenomenon can be ascribed to
properties of the medium formed in RHIC, which thus exhibits
a non-standard QFT behavior through the considered correlation functions.
Adopting that correlation function approaches $1+\lambda $ when
the relative momentum $\ {\bf p} \to \ 0$, we construct an effective model
capable to mimic real physical picture. For this, we use \cite{AGI}
$q$-{\it deformed} commutation relations and the techniques of
{\it q-boson statistics} which result in a partial supression
of the quantum statistical effects in many-particle systems.
%%%%%%%%%%%%%%%%%%%%%%%%%%%%%%%%%%%%%%%%%%

As expected, usage of appropriate $q$-algebra allows
to reduce the treatment of complex system of interacting particles to
that of a system of non-interacting ones, at the price of more
complicated (deformed) commutation relations.
The deformation parameter $q$ is viewed as an effective
(not universal) parameter which efficiently encapsulates most
essential features of complicated dynamics of the system under study.
For example, in the application of $q$-deformed
algebras to description \cite{Bo}
of rotational spectra of superdeformed nuclei, $q$
is also non-universal and assumes a particular value for each nucleus.
In the context of hadron theory the quantum
($q$-deformed) algebras also proved to be useful. Such a usage
significantly improves description of hadron characteristics both
in hadron scattering \cite{CA,Ch,JKM}
(nonlinearity of Regge trajectories) and in the sector of
hadron masses and mass sum rules \cite{Ga}.

%%%%%%%%%%%%%%%%%%%%%%%%%%%%%%%%%%%%
In this letter we propose to employ, for the system of pions
or kaons produced
in RHIC, the ideal $q$-Bose gas picture based on so-called $q$-bosons.
Physical meaning or explanation of the
origin of $q$-deformation in the considered phenomenon
depends on whether the deformation parameter $q$ is real or takes
complex value, say, a pure phase factor.
Here we do not deal with the diversity of all possible reasons
(not completely unrelated) for the appearance of $q$-deformed statistics.
Let us mention only that the {\it compositeness }
of particles (pions, kaons) may also lead \cite{GGG}
to a $q$-deformed structure with real $q$.

We use the set of $q$-oscillators defined as \cite{BM,AFZMP}:
\[
[N_i,b_j]=-\d_{ij} b_j\ ,\ \ \
[N_i,b^\dg_j]=\d_{ij} b^\dg_j\ ,\ \ \
[N_i, N_j]=0 \ ,
\]

\vspace{-4mm}
\[
[b_i,b_j]=[b^\dg_i,b^\dg_j]=0\ ,\ \ \ \
\]

\vspace{-3mm}
\beq
b_i b_j^\dg-q^{-\d_{ij}} b_j^\dg b_i=\d_{ij}q^{N_j} \ .   \label{a5}
\eeq
In the Fock type representation, with the ``$q$-bracket'' defined as
$[r]_q=(q^r-q^{-r})/(q-q^{-1})$, we have
\beq
b^\dg_i b_i=[N_i]_q \, .
\label{a6}
\eeq
The equality $b^\dg_i b_i = N_i$ holds only if $q=1$
(no-deforma\-tion limit). It is required that either $q$ is real or
\beq
q=\exp (i \t)\ ,
          \ \ \ \ \ \ \ \ \ \ 0 \le \t \le \pi \ .
\label{19}
\eeq
Below, just
this exponential form
will be adopted for $q$ (compare with the
phase $\alpha $ in Eqs.~(\ref{i5}),(3)).

For a multi-pion or multi-kaon system, we consider the model of
ideal gas of $q$-bosons (IQBG) taking the
Hamiltonian in the form \cite{AG,Go}
\beq
H=\sum_i{\o_i { N}_i}
\eeq

\vspace{-0.4cm}
\noindent with $i$ labeling energy eigenvalues,
$\omega_i=\sqrt{m^2+{\bf k}_i^2}$, and $N_i$ defined as above.
This is the unique truly noninter\-ac\-ting Hamiltonian
with additive spectrum. Also, we assume discrete
$3$-momenta of particles, i.e. the considered system is contained in
a large box of volume $\sim L^3$.

Basic statistical properties, as usual, are obtained by
evaluating thermal averages $ \la A \ra =
{\rm Sp}(A\rho )/{\rm Sp}(\rho)\ $, $\rho = e^{-\b H} , $
now with the Hamiltonian (9). Here $\b=1/T$ and
the Boltzmann constant is set equal to 1.

With $b^\dg_i b_i=[N_i]_q$ and $q+q^{-1}=[2]_q=2\cos\t$,
the $q$-deformed distribution function is obtained as
\beq
\la b_i^\dg b_i \ra=\frac{e^{\b\o_i}-1}
{e^{2\b\o_i}-2\cos(\theta)e^{\b\o_i}+1} \ .
\eeq
If $q\to 1$ it yields
the Planck-Bose-Einstein distribution, as
should, since at $q=1$ we return to the standard system of
bosonic commutation relations.
Although the deformation parameter $q$ is chosen in the
{\it complex} form (8), the
$q$-distribution function (10) turns out to be real, owing to
dependence on $q$ through the sum $q+q^{-1}$.

Note that the $q$-distribution (9) already appeared in refs. \cite{AG}
(moreover, a two-parameter generalization of distribution (10)
based on appropriate two-parameter deformed version
of the relations (6) is also possible \cite{DK}).

\vspace{-0.1cm}

%%%%%%%%%%%%%%%%%%%%%%%%%%%% fig1 %%%%%%%%%%%%%%%%%%%%%%%%%%%%
\begin{figure} [ht]
\begin{center}
\epsfig{file=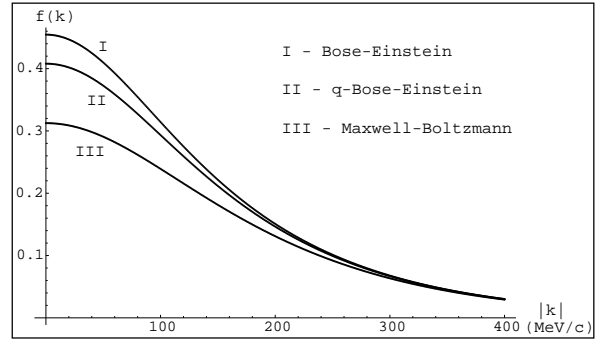,width=7.8cm,angle=0}
\vspace{0.6cm}
\caption{ The $q$-distribution function versus momentum (curve II),
in comparison with the quantum Bose-Einstein (curve I) and
Maxwell-Boltzmann (curve III) distributions. The inputs are:
$T = 120 $ MeV,  $m=m_{\pi} .$ Curve II corresponds to the
deformation angle $\t = 24^{\circ} $. }
\end{center}
\end{figure}
%%%%%%%%%%%%%%%%%%%%%%%%%%%%%%%%%%%%%%%%%%%%%%%%%%%%%%%%%%%%%

\vspace{-2mm}

The shape of the function $f(k)\equiv \la b^\dg b \ra (k) $
from (10) corresponding to the gas of pions modelled by IQBG is
given in Fig.~1. As seen,
the $q$-deformed distribution function lies completely
in between the well-known two curves (standard Bose-Einstein
distribution function and classical Maxwell-Boltzmann one), demonstrating
that the deviation of $q$-distribution (10) from the quantum
Bose-Einstein distribution goes in the ``right direction'' towards the
classical Maxwell-Boltzmann one, which reflects a decreasing of
the manifestation of quantum statistical effects.

Analogous curves for $q$-distribution functions
can be given with other fixed data.
For kaons, because of their larger mass
and higher empirical value $\l\simeq 0.88$ of the intercept
(which corresponds to a smaller deformation in our model),
such a curve should lie
closer (than pion's one) to that of the Bose-Einstein distribution.

Now let us go over to our main subject of two-particle correlations.
Calculation of the two-particle distribution
corresponding to the $q$-oscillators (\ref{a5}) yields
\beq
\la b_i^\dg b_i^\dg b_i b_i\ra=\frac{2\cos\theta}
{e^{2\b\o_i}-2\cos(2\theta)e^{\b\o_i}+1}\ .
\eeq
Then, the desired formula for the intercept
$\tilde\l_i\equiv\l_i + 1=
\la b_i^\dg b_i^\dg b_i b_i\ra / (\la b_i^\dg b_i \ra)^2$
of two-particle correlations follows,
\beq
\l_i + 1=
\frac{2\cos\theta(t_i+1-\cos\theta)^2}{t_i^2+2(1-\cos^2\theta)t_i}
\eeq
with $ t_i=\cosh(\b\o_i)-1$.
Note that (11), (12) are real functions since, like (10), 
these depend on the $q$-parameter (8) through the %particular 
combination $\frac12 (q+q^{-1})=\cos\t $.
Below, we drop the subscript $i$ and ignore the
above assumed discreteness of momenta.

The quantity $\l$ can be directly confronted with empirical data.
In the limit $q\to 1$, from Eq.~(12) the intercept $\l_{\rm BE}=1$
proper for Bose-Einstein statistics is reproduced. This
corresponds to Bose-Einstein distribution, see (10) at $\theta = 0$.
On the other hand, at $\theta=\pi/2$, Eq.~(12) coincides with
the value $\l_{\rm FD}=-1$ proper for Fermi-Dirac statistics.
The two cases, $\l_{\rm BE}$ and $\l_{\rm FD}$,
are seen in Fig.~2 as the only two
points where all the different curves (the continuum
parametrized by $w=\b \o$) merge and
the dependence on momentum and temperature disappears.
{}From the continuum of curves, there exists a 
unique limiting asymptotic one
$\tilde{\l}=2\cos\t $ (or $\l=-1+2\cos\t $),
which corresponds to $w\to\infty$, i.e. to low temperature or
to large momentum.

Let us discuss implications of Eq.~(12). Solving
it for $\cos\t$ at a fixed
$\l=\l_1$, we obtain the deformation angle as the function
$\t=\t(\l_1,\ {\bf K},\ T,\ m)$.
Fig.~2 illustrates the properties of intercept (correlation strength) $\l$
treated from the standpoint of $q$-deformation,
i.e., on the base of Eq.~(12).
Note that the continuum of curves $\tilde\l=\tilde\l (\cos\t )$
parametrized by $w=\b\o $ divides into three
classes (``subcontinua'') given by the
intervals:
(i) $0<w\le w_0 $, (ii) $w_0<w<w'_0 $, and (iii) $w'_0\le w<\infty $.
The two ``critical'' values $w_0=w_B\simeq 0.481\ $ and
$w'_0=w_D\simeq 0.696\ $
fix the curves B and D respectively.
The curves A, C, E are typical representatives of the
classes (i), (ii),  (iii). All the curves from classes (i), (ii)
possess two extrema;
the curve D is unique since its extrema degenerate, coinciding
with the point of inflection. This enables us to define
the range of ``small deformations''
$I_{\rm small}$ for the variable $\t$: from $\t =0$
(no deformation) to the value
yielding minimal $\l $, $\l_{\rm min}\approx 0.33$, implied by
the ``critical'' value $w'_0=w_D$.
On the interval $I_{\rm small}$ the intercept $\l $ monotonically
decreases with
increase of $\t$ (or $1-\cos\t $, the strength of deformation).

%%%%%%%%%%%%%%%%%%%%%%%%%%%%%%%% fig2 %%%%%%%%%%%%%%%%%%%%%%%%%%%%
\begin{figure}[t]
\begin{center}
\vspace{-0.01cm}
\epsfig{file=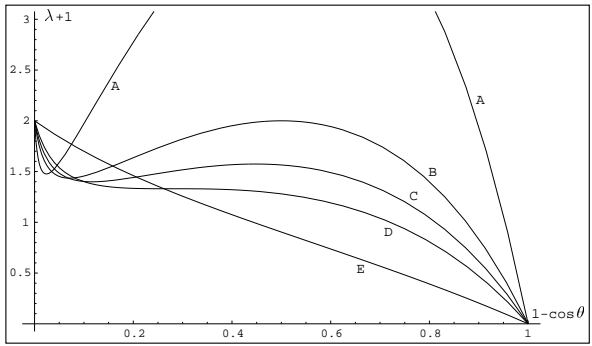,width=7.8cm,angle=0}
\vspace{0.1cm}
\caption  { The $(1-\cos\t)$-dependence of the
intercept $\l $ of two-particle correlation,
given by Eq.~(12). The curves A,B,C,D,E correspond to
the values $w_A= 0.3, $ $w_B=0.481, $ $w_C=0.58, $
$w_D=0.696, $ $w_E=2.0 $
of the dimensionless variable $w\equiv \b \o$.  }
\vspace{-0.2cm}
\end{center}
\end{figure}
%%%%%%%%%%%%%%%%%%%%%%%%%%%%%%%%%%%%%%%%%%%%%%%%%%%%%%%%%%%%%%%

All the type (ii)--(iii) curves at $\t\ne 0$ lie below the line
$\tilde\l = 2\ $ --- largest possible correlation
attainable in Bose-Einstein case
(the curve B contains, besides $\t =0 $,
{\it single} point at certain value of $1-\cos\t$, where
$\tilde\l = 2\ $ is also attained).
{}For the type (ii) curves we restrict ourselves with $I_{\rm small}$,
ignoring moderately (and very) large deformations.
The class (i) contains
``irregular'' curves for each of which there exist
$q$-deformations yielding
correlation strengths exceeding $\tilde\l = 2$.
Consider special values of physical variables
$T,\ |{\bf K}|$, which provide the peculiar values
$w_0=w_B\simeq 0.481 $ and  $w'_0=w_D\simeq 0.696 $ (recall that
$w=\sqrt{m^2 + {\bf K}^2}/T $). With $m(\pi^{\pm,0})=139.57$ MeV
and lowest mean momentum of the pion pair $|{\bf K}| = 0$, we get
the respective two lower bounds for the temperature: $T_0=290.0$~MeV
and $T'_0=200.5$~MeV.  Compare these values with
that for a typical curve from class (iii):
at $w=w_E=2.0 $ (the curve E) for pions with $|{\bf K}| = 0$
we get the lowest temperature $T_E=69.8$ MeV.

The results similar to presented above (starting with Eq.~(6))
were also given \cite{AGI} for $q$-oscillators with
defining relation $a a^\dagger - q a^\dagger a = 1$
introduced by Arik and Coon \cite{CA}.

Theoretical approaches to RHIC are aimed
to find an adequate description for the non-equilibrium state formed
during the collision,
and the $q$-boson technique enables us to treat the
non-stationary hot and dense matter effectively
as a ``noninteracting ideal gas''. To determine the $q$-parameter
corresponding to actual state of the hot medium,
we propose a way of extracting
a useful information from the two-particle correlations.
In a more general context,
by this we propose and develop an effective picture for %of
the two-particle correlations in RHIC.

Let us specially emphasize a remarkable feature
exhibited within our model:

{\it At $w\to\infty$
(i.e. for very low temperature at fixed momenta
or very large momenta at fixed temperature),
see Fig.~2 and relevant comments, we come to the equality}
\beq
   \hspace{1mm}  \l = 2\cos\t -1 \hspace{6mm}
      (T\to 0 \ \ {\rm or} \ \  \vert{\bf K}\vert \to\infty)
\eeq
{\it for the employed $q$-oscillators} (6). 
This means a direct link
$\l\leftrightarrow q $ whose explicit form is
$\tilde\l = [2]_q = 2 \cos\t$.

Finite temperature and momenta, on the other hand,
become non-trivially involved
in the relation between $\l$ and the parameter $q$ as seen from
Eq.~(12) and Fig.~2.

In Fig.~3, we present the dependence, implied by Eq.~(12),
of intercept $\lambda$ on the momentum $\vert{\bf K}\vert$
exemplifying this for pions at $T=120$ MeV (and at $T=180$ MeV)
with four curves
corresponding to the fixed values  $4^{\circ}$, $9^{\circ}$,
$15^{\circ}$, $24^{\circ}$ of the deformation angle $\theta$.
Analogous dependence for the case of kaons is presented in Fig. 4.
Each curve has its own asymptote given by Eq.~(13)
and lying beneath the line $\lambda =1$.
Recall that, if the unique cause for reducing of the intercept
would be decays of resonances (commonly accepted idea), all the
curves would tend to the value $\lambda =1$ in their
asymptotics.
In this respect, on the contrary,  
our model clearly suggests
possible existence of some other reasons,
than just the decays of resonances,
for the deviation of $\lambda$ from unity.
This property probably reflects complicated
dynamics of strongly interacting system prior to freeze-out.
The particular kind of the behaviour of $\lambda$ with respect
to $\vert{\bf K}\vert$ (implied for a fixed deformation),
is a direct consequence of our model.

%%%%%%%%%%%%%%%%%%%%%%%%%%%%%%%% fig3 %%%%%%%%%%%%%%%%%%%%%%%%%%%%
\begin{figure}[ht]
\begin{center}
\vspace{-0.01cm}
\epsfig{file=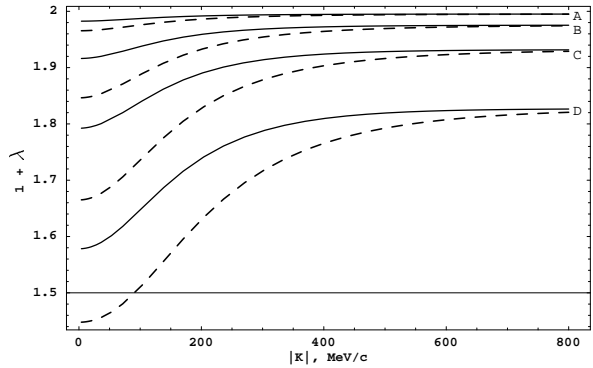, width=7.8cm ,angle=0}
\vspace{0.1cm}
\caption{The intercept $\lambda$ versus pion momentum $|{\bf K}|$ for
fixed $\theta$: A) $\theta = 4^\circ$, B) $\theta = 9^\circ$,
C) $\theta = 15^\circ$, D) $\theta = 24^\circ$.
The inputs are $T=120$ MeV - solid curve,
$T=180$ MeV - dashed curve, and $m=m_\pi$. }
\vspace{-0.2cm}
\end{center}
\end{figure}
%%%%%%%%%%%%%%%%%%%%%%%%%%%%%%%%%%%%%%%%%%%%%%%%%%%%%%%%%%%%%%%%%%

%%%%%%%%%%%%%%%%%%%%%%%%%%%%%%%% fig4 %%%%%%%%%%%%%%%%%%%%%%%%%%%%
\begin{figure}[ht]
\begin{center}
\vspace{-0.01cm}
\epsfig{file=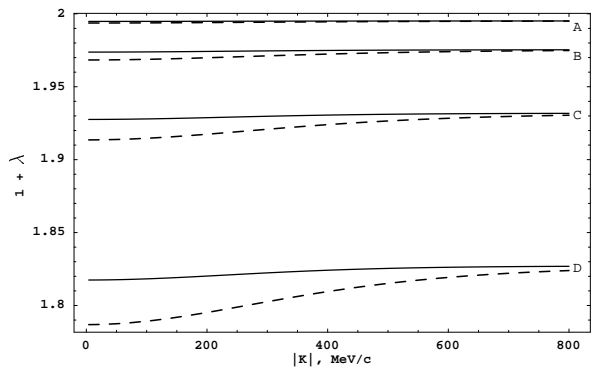, width=7.8cm ,angle=0}
\vspace{0.1cm}
\caption{The intercept $\lambda$ versus kaon  momentum
$|{\bf K}|$ for fixed $\theta$, with the same as in Fig. 3  values
of the parameters  $\theta$, $T$, and now $m=m_K$.
}
\vspace{-0.2cm}
\end{center} \end{figure}
%%%%%%%%%%%%%%%%%%%%%%%%%%%%%%%%%%%%%%%%%%%%%%%%%%%%%%%%%%%%%%%%%%

Let us summarize main points of the present letter.
In contrast to the commonly accepted opinion that
the intercept $\lambda$  of the two-pion correlations should
asymptotically approach the value one,
we predict attaining by  $\lambda$  the value less than
one at large pion (kaon) pair mean momentum $\vert{\bf K}\vert$.
This effect can be verified in the current experiments which take place
on the SPS at CERN and on the Relativistic Heavy Ion Collider at BNL.
Therefore we propose to pay special attention, in these experiments,
to the behaviour of the intercept $\lambda$
at large enough pion (kaon) pair mean momentum,
that is, $\vert{\bf K}\vert$ in the range
up to $500\!-\!600$ MeV/c for pions
(up to $700\!-\!800$ MeV/c for kaons) in the fireball frame.
If our prediction is confirmed, this means that $\lambda$
encapsulates certain memory effects:
it carries important information about the fireball dynamics,
especially concerning probable quark-gluon plasma phase, hadronization,
other nonstationary processes which took place in the fireball before
freeze-out.

%%%%%%%%%%%%%%%%%%%%%%%%%%%%%%%%%
As follows from our model, at large enough pair mean momenta of the two
pions or two kaons, the dependence of intercept $\lambda$ on temperature
as well as on the particle mass (and energy) is washed out.
Due to this, $\lambda_\pi$ and $\lambda_K$ should
practically coincide already at
$\vert{\bf K}\vert \simeq 800$ MeV/c,
although at much smaller mean momenta the intercepts
$\lambda_\pi$ and $\lambda_K$
significantly differ from one another (compare the corresponding
curves in Fig.~3 and Fig.~4, labelled by the same $\theta$).
Let us recall that this remarkable feature is the result of asymptotical
reducing of the full $T$-, ${\bf K}$-, and $m$-dependence contained in
Eq.~(12) to the direct link between the intercept $\lambda$
and the deformation parameter $q\ $ (with $ q=\exp{{\rm i}\theta}$),
given by Eq.~(13).
Therefore, {\em we put forward two principal points
for experimental verification}:\\
%$\hspace{3.6mm}$ {\bf 1.}
1) The shape of dependence of intercept $\lambda$ on
the pair mean momentum as well as
its asymptotics (constant value of
$\lambda_\pi$, $\lambda_K$ less than unity at large
$\vert{\bf K}\vert$);\\
%$\hspace{3.6mm}$ {\bf 2.}
2) Coincidence of the intercepts for different
pairs of particles at
large enough mean momentum, for instance, the regime with
$\lambda_\pi \approx \lambda_K$ should manifest itself starting
%practically
already from %the mean momentum value
$\vert{\bf K}\vert \simeq 800$ MeV/c.

Verification of these issues is of fundamental interest.
Indeed, the first point will give answer to the question:
is the q-deformed description ($q$-boson field theory) really capable
to mimic the unusual properties of "hot" and "dense" hadron matter?
In particular, can we adopt this description as an efficient tool
to deal with nonequilibrium, short-lived system of pions or kaons at
high temperature and density?
The second point of verification will clarify %the question
whether one can regard the deformation parameter $q$
as a kind of quantity, which may be referred to as one of 
basic characteristics of the hot, short-lived system
(besides "temperature" and chemical potential).

We hope that the answers to these intriguing questions will be deduced
in the nearest future in the experiments which take place in CERN and BNL.

%%%%%%%%%%%%%%%%%%%%%%%%%%%%%%%%%%%%%%%%%%%%%%%
D.A. acknowledges stimulating discussions with U.~Heinz and valuable
advise of P.~Braun-Munzinger to present the picture in the case of
kaons (Fig.~4).
A.G. thanks V.~Man'ko and F.~Zaccaria for the discussion of results.
The work of A.G. and N.I. was partially supported by
the Award No. UP1-2115 of the U.S. Civilian Research and
Development Foundation (CRDF).

%%%%%%%%%%%%%%%%%%%%%%%%%%%%% bibliography %%%%%%%%%%%%%%%%%%%%%%%%%%

\end{document}